\documentclass[11pt,a4paper,notitlepage]{article}
\usepackage{titling}
\pdfoutput=1
\usepackage[utf8]{inputenc}
\usepackage{authblk}
\usepackage[english]{babel}
\usepackage{amsmath}
\usepackage{amsfonts}
\usepackage{amssymb}
\usepackage{graphicx}
\usepackage{epstopdf}
\usepackage[english]{babel}
\usepackage[autostyle, english=american]{csquotes}
\MakeOuterQuote{"}
\title{Anticipation: an effective evolutionary strategy for a sub-optimal population in a cyclic environment}
\author[1]{D. V. Arjun\thanks{darjun@ibab.ac.in}}
\author[2]{Pallab Basu\thanks{pallab.basu@icts.res.in}}
\author[1]{Sayak Mukherjee\thanks{Corresponding Author: vilagrussa@gmail.com}}
\affil[1]{Institute of Bioinformatics \& Applied Biotechnology, Biotech Park, Electronic City, Bangalore 560100, India}
\affil[2]{International Centre for Theoretical Sciences, Tata Institute of Fundamental Research, Bangalore, 560089, India}
\begin{document}
\date{}
\maketitle

\begin{abstract}
We built a two-state model of an asexually reproducing organism in a periodic environment endowed with the capability to anticipate an upcoming environmental change and undergo pre-emptive switching. By virtue of these anticipatory transitions, the organism oscillates between its two states that is a time $\theta$ out of sync with the environmental oscillation. We show that an anticipation-capable organism increases its long-term fitness over an organism that oscillates in-sync with the environment, provided $\theta$ does not exceed a threshold. We also show that the long-term fitness is maximized for an optimal anticipation time that decreases approximately as $1/n$, $n$ being the number of cell divisions in time $T$. Furthermore, we demonstrate that optimal "anticipators" outperforms "bet-hedgers" in the range of parameters considered. For a sub-optimal ensemble of anticipators, anticipation performs better to bet-hedging only when the variance in anticipation is small compared to the mean and the rate of pre-emptive transition is high. Taken together, our work suggests that anticipation increases overall fitness of an organism in a periodic environment and it is a viable alternative to bet-hedging provided the error in anticipation is small. 
\end{abstract}
The question of how to best navigate an unpredictable future is central to the issue of survival of a species. It is widely believed that an evolutionary trade-off between immediate and long-term gain is essential to safeguard the overall success of a species. A decision to forego immediate gain is justified provided that the same decision effectively shields an organism from long-term risks. An annual plant for example, can selectively delay germination, thereby suffering an immediate loss in terms of its reproductive success, in order to avoid germinating all at once and accidentally facing a drought \cite{cohen1966optimizing}, \cite{cohen1967optimizing}. In this context of risk-gain trade-off, two survival strategies have been widely recognized: i) a {\it{bet-hedging}} strategy of maintaining phenotypic heterogeneity, ii) a strategy of  {\it{anticipation}} where an organism undergoes pre-emptive change in advance in order to address future problems. Using a well-known two state model, we explore the role of anticipation in maximizing evolutionary fitness of an asexually reproducing organism in a periodic environment and also conduct a comparative study of the same with a bet hedging model.  

Biology is replete with examples of bet-hedging, where a clonal population maintains non-heritable phenotypic polymorphism even though only one of the phenotypes is fit in a given environment. It was known since 1942 for example, that a fraction of bacteria that were not resistant, {\it{persisted}} nonetheless even after extensive treatment with antibiotic \cite{hobby1942observations}, \cite{bioger1944treatment}.  The persisters were shown to be slow-growing or dormant \cite{balaban2004bacterial} and hence could not be targeted by antibiotics. Under normal condition, persisters, owing to their slow growth, were evidently unfit, yet a bacterial population always maintained them as an "insurance" against antibiotic stress. Phenotypic polymorphism is fairly common in systems governed by non-linear gene regulatory mechanisms \cite{arkin1998stochastic},\cite{novick1957enzyme},\cite{acar2005enhancement},\cite{gardner2000construction},\cite{isaacs2003prediction},\cite{ozbudak2004multistability} where a globally stable state splits into two or more stable states as a function of an external parameter \cite{Strogatz1994}. E. Coli cells for example, cultured in a lac non-induced state split into two stable sub-groups, one fully induced and another totally non-induced, as a function of external inducer concentration \cite{ozbudak2004multistability}. Another salient feature of these types of systems is spontaneous stochastic phenotypic switching between phenotypes as exemplified in bacterial persistence \cite{balaban2004bacterial}. Using {\it{hipA7}} and {\it{hipQ}} mutants of E. Coli, Balaban et al. identified two different types of persisters, which they named Type I persister and Type II persister, along with the normal rapidly dividing cells. They showed that while Type I persisters were characterized by a negligible switching from normal to the persistent phenotype during exponentially growing phase, switching could be induced by trigger events like antibiotic stress. Type II persisters on the other hand had non-zero switching rates between the normal and persister phenotypes regardless of trigger events. Therefore, it is clear that these switching rates are not only essential in maintaining phenotypic diversity but they are themselves influenced by environmental trigger events \cite{acar2008stochastic}. 

While bet-hedging has been established as a ubiquitously deployed strategy for maximizing long-term fitness, it is not the only currency in town. Another widely discussed strategy is anticipation wherein an organism anticipates an impending change and mounts a pre-emptive response. To quote Rosen, "An anticipatory system is a system containing a predictive model of itself and/or its environment, which allows it to change state at an instant in accord with the model's predictions pertaining to a later instant" \cite{rosen2012anticipatory}. While it is difficult to conceive of a system that can anticipate a randomly fluctuating environment, it is unclear if environments in general are really given to vagaries. Examples of strictly periodic environments, like the seasonal or the diurnal cycles, are pretty commonplace. For an organism occupying a niche that changes cyclically, anticipatory behavior can be evolutionary hardwired. In line with this logic we see widespread anticipatory behavior in seasonal flowering plants and in migratory behavior of birds and mammals. Circadian oscillation is observed in mammals like mice and birds \cite{yamazaki2000resetting},\cite{deguchi1979circadian}, fruitfly\textemdash Drosophila melanogaster \cite{plautz1997independent},\cite{giebultowicz2000transplanted}, filamentous fungus\textemdash Neurospora crassa \cite{lakin2004circadian}, and also in cyanobacterium\textemdash Synechococcus elongatus \cite{cohen2015circadian}. Defining property of the oscillator is its ability to be entrained by an external time varying stimulus and a free running rhythm in the absence of external reinforcement. Free running rhythm constitutes a perfect example of anticipatory behavior. Furthermore, as correctly pointed out by Tagkopoulos et al. \cite{tagkopoulos2008predictive}, fluctuation in one environment variable can be highly correlated with others.  Temporal fluctuation in temperature can appear to be random but when taken together with fluctuation in $O_{2}$ levels, can show highly correlated structure: a rise in the former heralding a drop in the later \cite{tagkopoulos2008predictive}. Using a combination of computer simulation and experimental validation, the authors showed that these temporal correlations could be exploited even by simple microbes to mount an anticipatory response. Given the obvious fitness benefit in accurately predicting changes in an environment, it stands to reason to study anticipation quantitatively using analytically tractable models. Previously, Kussell et al. \cite{kussell2005phenotypic} undertook a similar venture by building a linear model and comparing two strategies: bet hedging and what they called "responsive-switching"\textemdash a mechanism akin to anticipation. Their "responsive-switching" was incorporated in the structure of the switching matrix and also in a cost variable $c$, representing the cost for maintaining an active sensory device. Assuming that the duration of different environmental states are long, they showed that stochastic switching is favored when environment stays steady for longer. Though their model accounted for the cost of maintaining a sensory apparatus, the phenotypic switching rates however changed instantaneously, without delay and in sync with the environment. Therefore their model did not explicitly take into account pre-emptive response\textemdash a major tenet of anticipatory behavior. 

Two state models have a long history in mathematical evolutionary biology \cite{balaban2004bacterial},\cite{thattai2004stochastic},\cite{kussell2005bacterial},\cite{lachmann1996inheritance}. In this paper, we modified an existing two state model discussed in Thattai et al. \cite{thattai2004stochastic} and made it anticipation capable by explicitly endowing it with a capacity to mount a pre-emptive response and then answered questions like\textemdash can anticipation confer fitness benefit? Second, what is the optimal anticipation time and range of time over which anticipation is beneficial? What are the factors these quantities depend on and how? For a strictly cyclic environment, does anticipation fare better to bet hedging? How does error in anticipation affect fitness? 

\section*{Model}

\begin{figure}
\begin{center} 
\includegraphics[width=12cm,height=10cm]{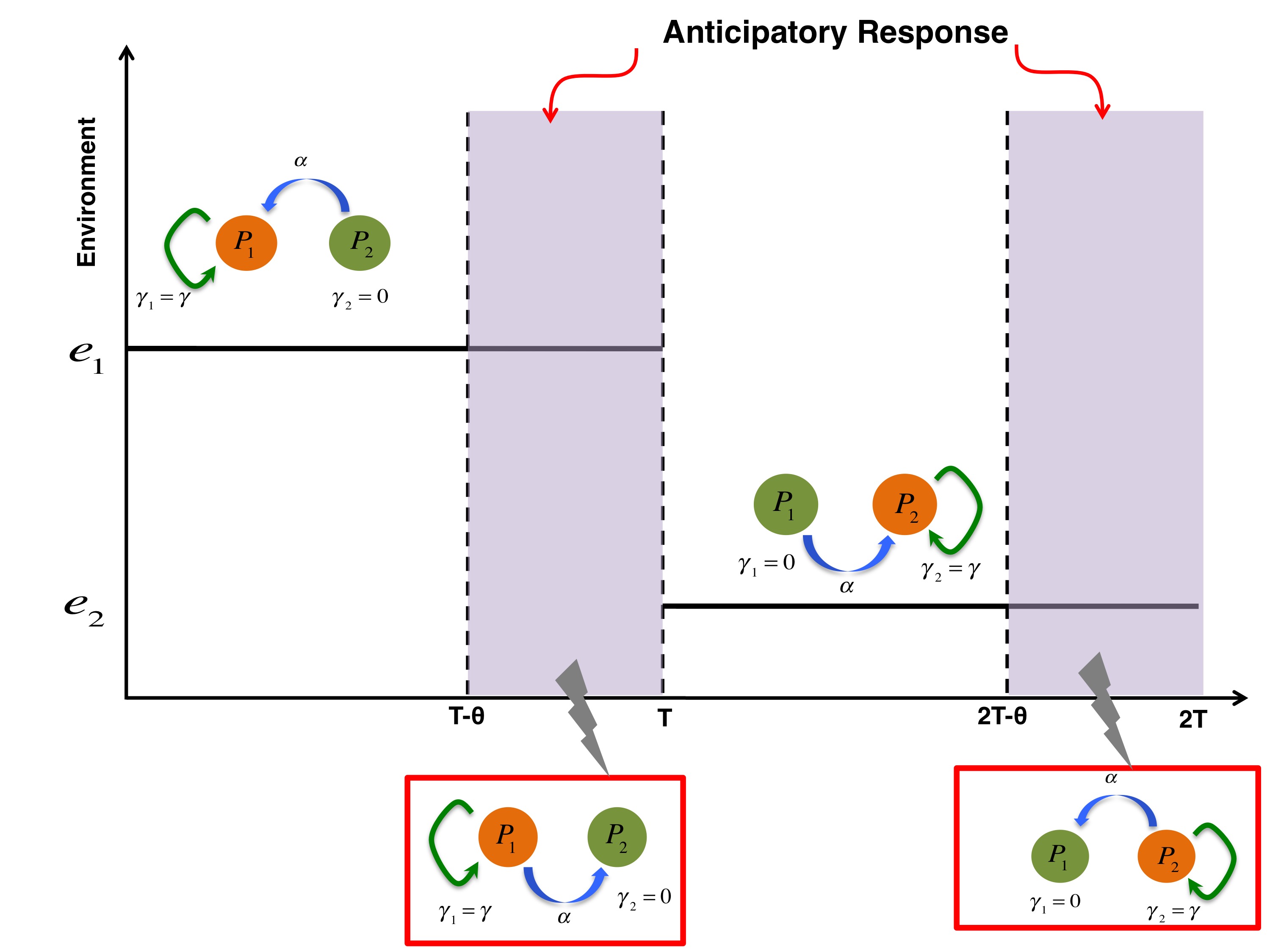}
\caption{We assume a periodically changing environment that shuttles between two environmental states $e_{1}$ \& $e_{2}$, shown as black horizontal bars, with a period $2T$. The organism has two phenotypic states, $P_{1}$ and $P_{2}$, color-coded according to their fitness. The fit phenotype is shown in orange and the unfit phenotype in green. The fit phenotype grows with a rate $\gamma>0$ (shown as a self loop in green) whilst the unfit phenotype is dormant and does not grow. Our organism surveys its environment and by virtue of its ability to monitor the environment, phenotypic switchings are not random. So in $e_{1}$, the unfit phenotype $P_{2}$ switches to $P_{1}$ with a rate $\alpha$ for a time period of $T-\theta$. In an anticipation for an upcoming environmental flip, the organism mounts a pre-emptive response by reversing the direction of transition by setting the rate from $P_{1}\rightarrow P_{2}$ to $\alpha$ and from $P_{2}\rightarrow P_{1}$ to zero, at a time $T-\theta$ and continues like that until $2T-\theta$ when it mounts its second anticipatory response. At a time $T$ the environment switches from $e_{1}$ to $e_{2}$, rendering $P_{1}$ unfit and $P_{2}$ fit. $P_{2}$ therefore proliferates with a rate $\gamma>0$ but $P_{1}$ now lays dormant. At $2T-\theta$, the organism reverses the direction of transition once more before the environment switches back to $e_{1}$ from $e_{2}$ at time $2T$. The shaded regions from $T-\theta$ to $T$ and from $2T-\theta$ to $2T$ are anticipatory regimes where a phenotype, in spite of being fit, pre-emptively transitions to the unfit phenotype. The modules used for anticipatory regimes are enclosed inside red rectangular boxes.} 
\label{Model}
\end{center}
\end{figure}

Our model is shown graphically in {\bf{Fig \ref{Model}}}. It describes evolution of a population of asexually reproducing organism in a periodically changing environment that shuttles between two states, $e_{1}$ and $e_{2}$, with a period $2T$. We assumed that our model organism has two distinct phenotypic states, denoted by $P_{1}$ and $P_{2}$, with $P_{1}$ being fit in $e_{1}$ and $P_{2}$ in $e_{2}$ respectively. Therefore, our organism exchanges fitness between phenotypes $P_{1}$ and $P_{2}$ as the environment cycles between $e_{1}$ and $e_{2}$. Fitness is measured in terms of their growth rates, $\gamma_{1}$ and $\gamma_{2}$, with the fit phenotype having a higher growth rate. We set the growth rate of the unfit phenotype to be zero ($\gamma_{2}=0$ in $e_{1}$ and $\gamma_{1}=0$ in $e_{2}$), thereby tacitly assuming that the unfit phenotype is dormant. In principle, fit phenotypes \textemdash $P_{1}$ in $e_{1}$ and $P_{2}$ in $e_{2}$ \textemdash can have two different growth rates in these two different environments but for simplicity's sake we have assumed them to be the same and denoted it by $\gamma$. Main conclusions of this paper are unaffected by this assumption. It is further assumed that the organism can anticipate an impending environmental transition at a time $\theta$ before the environment actually flips by means of a sensory apparatus that constantly surveys the environment. By virtue of this assumption, phenotypic switchings in our model are not random but instead they are regulated actively by the organism itself. Therefore, for $e_{1}$, only the unfit phenotype $P_{2}$ transitions to $P_{1}$ with a rate $\alpha$ but  $P_{1}$ is forbidden to change to $P_{2}$. Had it been random, transitions from both fit to unfit and from unfit to fit phenotypes would have been non-zero. After a time $T-\theta$ has elapsed, our organism senses an upcoming environmental change and subsequently switches its rates pre-emptively whereby the rate of transition from $P_{1}\rightarrow P_{2}$ becomes $\alpha$ and the transition rate from $P_{2}\rightarrow P_{1}$ becomes zero. It is worth noting that for $T-\theta\le t\le T$, $P_{1}$ is still the fit phenotype ($\gamma_{1}=\gamma$ and $\gamma_{2}=0$) yet the organism bears this immediate cost of switching to the unfit phenotype in an effort to prepare for the upcoming changes in the future. After a time $T$, the environment changes from $e_{1}$ to $e_{2}$ rendering $P_{2}$ fit over $P_{1}$ and consequently $\gamma_{2}$ becomes $\gamma$ and $\gamma_{1}$ becomes zero. At a time $2T-\theta$\textemdash a time $\theta$ before the second environmental flip\textemdash our model organism mounts its second pre-emptive response by resetting the transition rate from $P_{2}\rightarrow P_{1}$ back to $\alpha$ and from $P_{1}\rightarrow P_{2}$ to zero even though  $P_{2}$ is fit compared to $P_{1}$ in the time interval $2T-\theta\le t\le 2T$. After a time $2T$, the environment changes again from $e_{2}$ to $e_{1}$ and then the cycle repeats itself. 
Re-defining the system in terms of dimensionless parameters $\tau=\gamma t$, $\alpha=\alpha/\gamma$ we have
\begin{eqnarray}
  \left.
  \begin{aligned}
    N_{1,\tau+1}&\;=\;2N_{1,\tau}+\alpha N_{2,\tau}\nonumber\\
    N_{2,\tau+1}&\;=\;N_{2,\tau}-\alpha N_{2,\tau}
  \end{aligned}
  \right\}0\leq \tau\leq n-m\\
\nonumber\\
\left.
  \begin{aligned}
    N_{1,\tau+1}&\;=\;2N_{1,\tau}-\alpha N_{1,\tau}\nonumber\\
    N_{2,\tau+1}&\;=\;N_{2,\tau}+\alpha N_{1,\tau}
  \end{aligned}
  \right\}n-m\leq \tau\leq n\\
\nonumber\\
\left.
  \begin{aligned}
    N_{1,\tau+1}&\;=\;N_{1,\tau}-\alpha N_{1,\tau}\nonumber\\
    N_{2,\tau+1}&\;=\;2N_{2,\tau}+\alpha N_{1,\tau}
  \end{aligned}
  \right\}n\leq \tau\leq 2n-m\\
\nonumber\\
\left.
  \begin{aligned}
    N_{1,\tau+1}&\;=\;N_{1,\tau}+\alpha N_{2,\tau}\nonumber\\
    N_{2,\tau+1}&\;=\;2N_{2,\tau}-\alpha N_{2,\tau}
  \end{aligned}
  \right\}2n-m\leq \tau\leq 2n\\\label{Difference Equation}
\end{eqnarray}
where $N_{1}$ and $N_{2}$ are the sizes of phenotypes $P_{1}$ and $P_{2}$ respectively. It is also understood that the increment in time, $\Delta t$, is measured in units of cells division time scale $t_{d}=\gamma^{-1}$. Consequently $n=\gamma T$ and $m=\gamma\theta$ are the number of cell division events in times $T$ and $\theta$ respectively. The time $T$ is chosen in such a way that it is larger than a typical cell division time scale yet not large enough that the organism has ample time to adapt. In fact one can expect that in the limit of very large $T$, the advantage from anticipation will be negligible. From now on we will use $n$ and $m$ to indicate $T$ and $\theta$ respectively. The dimesionless rate $\alpha$ is the ratio of the actual transition rate to the growth rate and hence only $\alpha$ values of $<1$ are considered. It is clear from Eqn (\ref{Difference Equation}) that the number of individuals after $C$ cycles of period $2n$ is given by
\begin{eqnarray}
\vec{N}_{C}&=&\left[\begin{pmatrix} 1 & \alpha \\ 0 & 2-\alpha \end{pmatrix}^{m}\begin{pmatrix} 1-\alpha & 0 \\ \alpha & 2 \end{pmatrix}^{n-m}\begin{pmatrix} 2-\alpha & 0 \\ \alpha & 1 \end{pmatrix}^{m}\begin{pmatrix} 2 & \alpha \\ 0 & 1-\alpha \end{pmatrix}^{n-m}\right]^{C}\vec{N}_{0}\nonumber\\
&=&\left[\mathbb{M}_{4}^{m}\mathbb{M}_{3}^{n-m}\mathbb{M}_{2}^{m}\mathbb{M}_{1}^{n-m}\right]^{C}\vec{N}_{0}\label{Final State}
\end{eqnarray}
where $\vec{N}=(N_{1},N_{2})^{\rm{T}}$. Long term fitness therefore is given by the largest eigenvalue ($\lambda_{1}$) of the transfer matrix $\mathbb{M}_{4}^{m}\mathbb{M}_{3}^{n-m}\mathbb{M}_{2}^{m}\mathbb{M}_{1}^{n-m}$ \cite{kussell2005bacterial},\cite{lachmann1996inheritance}.
\section*{Results}
\begin{figure}
\begin{center} 
\includegraphics[width=10cm,height=8cm]{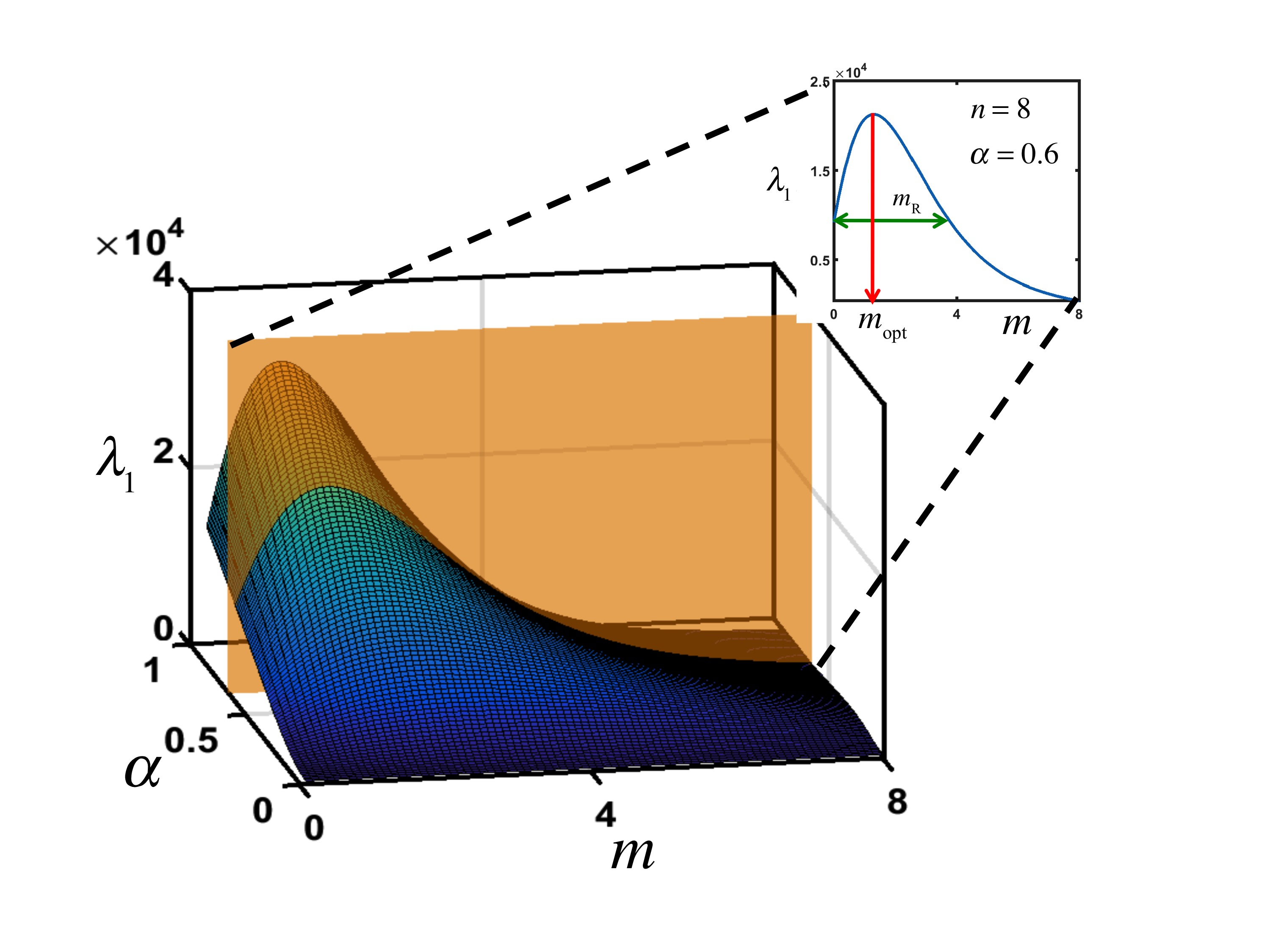}
\caption{Surface plot of the largest eigenvalue ($\lambda_{1}$) of the transfer matrix in Eqn (\ref{Final State}) as a function of the phenotypic switching rate $\alpha$ and anticipation time $m$ when $n=8$. The inset shows the variation of $\lambda_{1}$ as a function of $m$ for $\alpha=0.6$. The time when $\lambda_{1}$ reaches its maximum is denoted as $m_{\rm{opt}}$ and is shown with the red arrow. The range $m_{R}$ is shown with a double headed green arrow.} 
\label{Eigen surface}
\end{center}
\end{figure}

{\it{Anticipation increases long-term fitness}}: {\bf{Fig \ref{Eigen surface}}} shows the plot of long-term fitness ($\lambda_{1}$) as a function of $\alpha$ and $m$ for $n=8$. We see that for any $\alpha$, increase in $m$ leads to a transient increase in $\lambda_{1}$ reaching a maximum for an optimal $m$ ($m_{\rm{opt}}$). Beyond $m_{\rm{opt}}$, $\lambda_{1}$ decreases monotonically, crossing $\lambda_{1}(m=0)$ at $m=m_{R}$. From $m_{R}$ onwards, $\lambda_{1}$ progressively falls behind $\lambda_{1}(m=0)$ with increase in $m$, leading to a loss in long-term fitness compared to $m=0$ (no anticipation) case. With a decrease in $\alpha$, $\lambda_{1}(m_{\rm{opt}})$ decreases but the range $m_{R}$ over which anticipation pays off increases. For a slice at $\alpha=0.6$ for instance, given by the intersection of the $\lambda_{1}$ and the orange vertical plane (shown in the inset), $\lambda_{1}$ reaches its maximum ($\lambda_{1}^{\rm{max}}=2.13\times10^{4}$) for $m_{\rm{opt}}=1.3$. Beyond $m_{\rm{opt}}=1.3$, $\lambda_{1}$ decreases monotonically, stooping below $\lambda_{1}(m=0)=9.21\times10^{3}$ value at $m=m_{R}=3.81$. Using any other values of $n$ does not change any qualitative features of {\bf{Fig \ref{Eigen surface}}}. Therefore, it can be concluded that at least for a strictly periodic environment, the ability of an organism to anticipate change in advance ($m\neq 0$) confers long-term fitness advantage provided the anticipation time does not exceed a certain value $m_{R}$. Moreover with a decrease in $\alpha$, $m_{R}$ increases, enhancing the feasibility of anticipation.

\begin{figure}
\begin{center} 
\includegraphics[width=12cm,height=8cm]{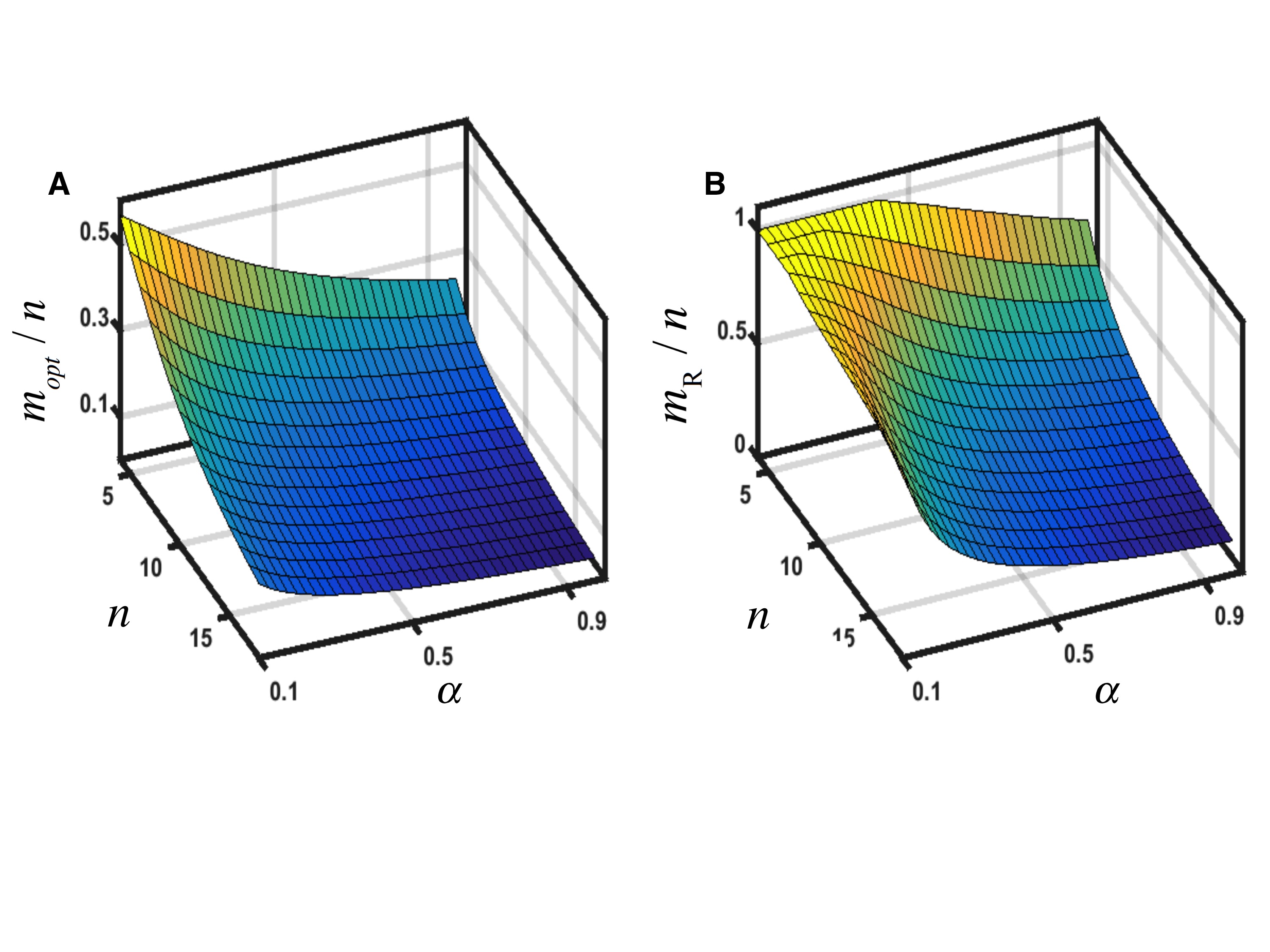}
\caption{{\bf{(A)}} A surface plot of $m_{\rm{opt}}/n$ as a function of $n$ and $\alpha$. {\bf{(B)}} A surface plot of $m_{R}/n$ as a function of $n$ and $\alpha$. $m_{R}/n=1$ is set by hand for those values of $n$ and $\alpha$ for which $m_{R}>n$.} 
\label{m optimal}
\end{center}
\end{figure}

In {\bf{Fig \ref{m optimal}}}, we show how $m_{\rm{opt}}/n$ and $m_{R}/n$ vary as functions of $n$ and $\alpha$. We find that $m_{\rm{opt}}/n$ decreases monotonically with increase in $n$ and $\alpha$ ({\bf{Fig \ref{m optimal}}A}). A trite calculation shows that $m_{\rm{opt}}/n$ decreases approximately as $1/n$ for moderate to high values of $\alpha$ (see Appendix A1, {\bf{Fig \ref{m_best all plots}}}, first two panels). For small values of $\alpha$, we see a $1/n$ decay in $m_{\rm{opt}}/n$ over a range of $n$, but with growing $n$, $m_{\rm{opt}}/n$ starts to decrease in a non-$1/n$ fashion (see {\bf{Fig \ref{m_best all plots}}}, third panel). This largely inverse relationship with $n$ is to be expected because as the number of cell division events increase, an organism gets more time to adapt which renders anticipation unnecessary as indicated by a decrease in $m_{\rm{opt}}/n$.  Similarly, as $\alpha$ goes up, we expect that the necessity for anticipation should go down as well because one can then respond to an environmental change by quickly switching from a non-advantageous phenotype to the fit phenotype and it does not need to anticipate in advance. In keeping with this logic, we also find an almost similar behavior in $m_{R}/n$ ({\bf{Fig \ref{m optimal}}B}) as well. The flat top in {\bf{Fig \ref{m optimal}}B} is due to the fact that $m_{R} > n$ for low values of $n$ or $\alpha$, implying that anticipation confers fitness advantage over the entire range of $m\in\left[0,n\right]$. For those cases, we have set $m_{R}/n=1$ by hand. For rest of the values of $n$ and $\alpha$, when $m_{R}<n$, we see that $m_{R}$ indeed decreases with increase in $n$ and $\alpha$. A semi-analytical approximation to $m_{R}/n$ can be obtained for large $\alpha$ values (see Appendix A1, {\bf{Fig \ref{m_range all plots}}}).

In sum, for a strictly periodic environment, anticipation pays off provided the environment does not stay unchanged for a long period of time or the inter-phenotypic transition rate is not too high driving $m_{\rm{opt}}/n$ and $m_{R}/n$ closer and closer to zero. 

\begin{figure}
\begin{center} 
\includegraphics[width=12cm,height=10cm]{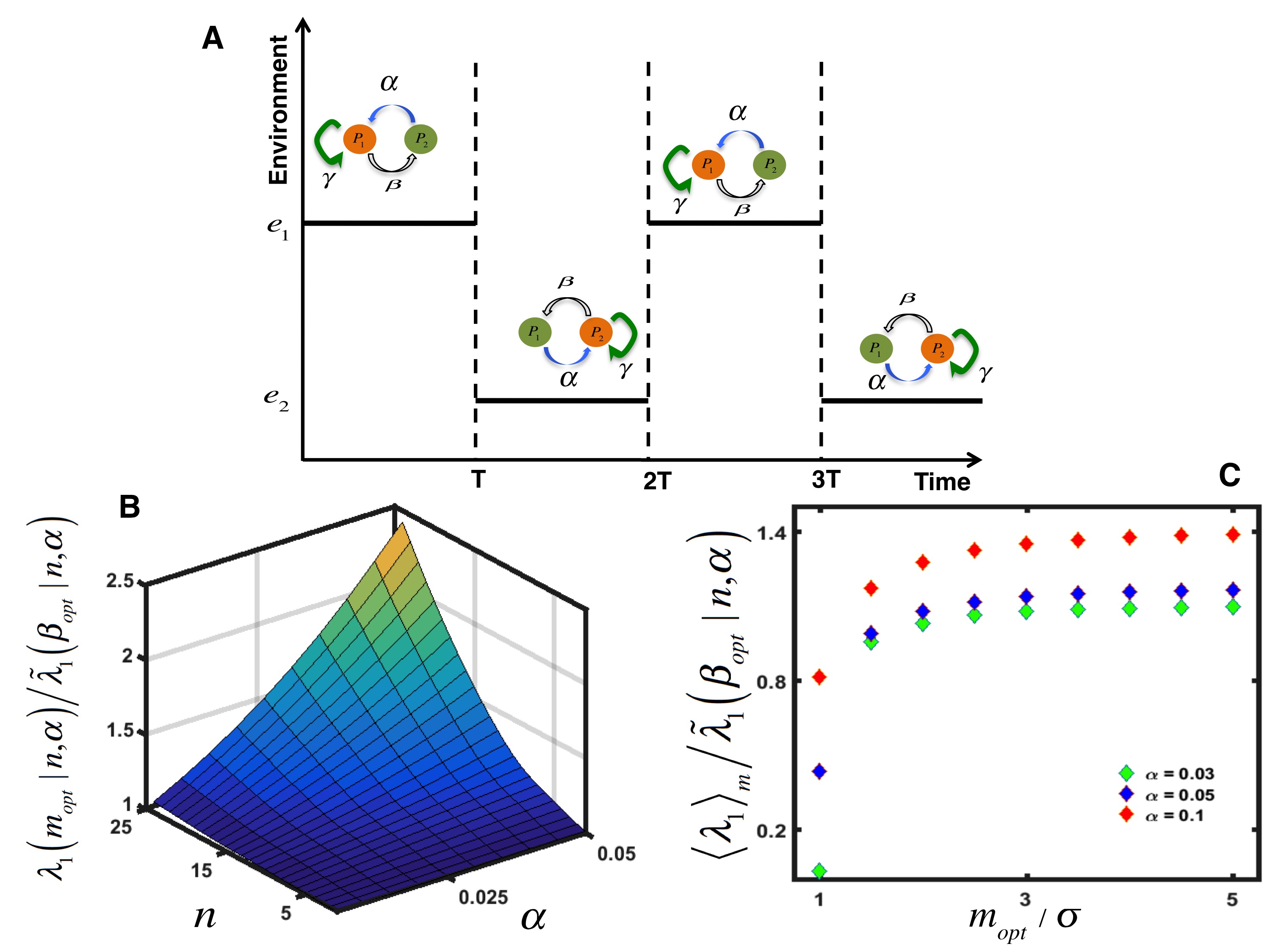}
\caption{{\bf{(A)}} Two state bet-hedging model of Thattai et al. The two phenotypes $P_{1}$ and $P_{2}$ can inter-convert. The unfit phenotype (green) switches to the fit phenotype (orange) with a rate $\alpha$ whereas the fit phenotype switches to the unfit phenotype with a rate $\beta<\alpha$. Fit phenotypes reproduces asexually at a rate $\gamma$ whereas the unfit phenotype is assumed to be dormant. The phenotypes exchange their fitness and switching rates synchronously with the environment. {\bf{(B)}} Plot of $\lambda_{1}(m_{\rm{opt}}|n,\alpha)/\tilde{\lambda}_{1}(\beta_{\rm{opt}}|n,\alpha)$ as a function of $\alpha$ and $n$.  {\bf{(C)}} Plot of $\langle\lambda_{1}\rangle_{m}/\tilde{\lambda}_{1}\left(\beta_{opt}|n,\alpha\right)$ as a function of signal to noise ratio $m_{\rm{opt}}/\sigma$ for three different $\alpha$ values and for $n=8$. Samples were drawn from Beta distribution with $\bar{m}=m_{\rm{opt}}$. The symbol $\langle ... \rangle_{m}$ stands for an ensemble average over all possible $m$ values.} 
\label{bet hedging}
\end{center}
\end{figure}

\vspace{0.5cm}
{\it{Anticipation is superior to bet hedging}}: Thattai et al. proposed a two state model of bet hedging in an asexually reproducing population \textemdash a model that inspired this work \cite{thattai2004stochastic}. In their model, they did not incorporate anticipation: $m$ was set to zero. Instead, they investigated the long-term fitness advantage of maintaining phenotypic heterogeneity. Consequently, they had a non-zero switching rate $\beta < \alpha$ from fit to the unfit phenotype along with a non-zero $\alpha$ and these switching rates reversed their values synchronously with the environment ({\bf{Fig \ref{bet hedging}}A}). Like before, the long-term fitness in their model is given by the largest eigenvalue ($\tilde{\lambda}_{1}$) of the transfer matrix $\tilde{\mathbb{M}}_{3}^{n}\tilde{\mathbb{M}}_{1}^{n}$, where $\tilde{\mathbb{M}}_{1}=\begin{pmatrix} 2-\beta & \alpha \\ \beta & 1-\alpha \end{pmatrix}$ and $\tilde{\mathbb{M}}_{3}=\begin{pmatrix} 1-\alpha & \beta \\ \alpha & 2-\beta  \end{pmatrix}$. They showed that for a periodic environment, $\beta\ne 0$ leads to an increase in long-term fitness compared to $\beta=0$ case, provided $\alpha$ value was not too large. Since the parameter $\beta$ serves to heterogenize a population into distinct phenotypes even for a constant environment; they concluded that heterogeneity increases long-term fitness. 

While our model incurs a short term cost in pre-emptive transition from fit to unfit phenotype at a time $m$ before the environment changes (shaded regions in {\bf{Fig \ref{Model}}}), their model also bears a short term cost by maintaining a sub population that is unfit. Therefore, it is not a priori clear which one of these two models would yield a higher long-term fitness for same values of $\alpha$ and $n$.  {\bf{Fig \ref{bet hedging}}B} shows the ratio of $\lambda_{1}(m_{\rm{opt}}|n,\alpha)$ to $\tilde{\lambda}_{1}(\beta_{\rm{opt}}|n,\alpha)$, where $\beta_{\rm{opt}}$, is the optimal transition rate  from an fit to an unfit phenotype for which the  $\tilde{\lambda}_{1}$ attains a maximum for a given $\alpha$ and $n$. We purposefully chose small values of $\alpha$ as phenotypic transitions cost energy and therefore small values of $\alpha$ are perhaps most relevant for our current discussion. We find that over the range of $n$ and $\alpha$ considered, our model consistently out-performed bet-hedging as is indicated by a ratio that is always greater than $1$. 

Next we considered an ensemble of {\it{anticipators}}. Each anticipator, characterized by an  $m$ value, was drawn from a Beta-distribution $f(m, a,b)=\left(m/n\right)^{a-1}\left(1-m/n\right)^{b-1}/\left(nB(a,b)\right)$, where $B(a,b)$ is the beta function, such that the ensemble average $\bar{m}=\int mf(m,a,b)dm=m_{\rm{opt}}$.  This ensemble can be thought of as a sub-optimal population of anticipators, each of whom makes an error $e=\mid m-m_{\rm{opt}}\mid$ in its anticipation, but the population average of the error still sums to zero. The long-term fitness is given by the ensemble average $\langle\lambda_{1}\rangle_{m}= \int\lambda_{1}(m|n,\alpha)f(m,a,b)dm$ where $a$ and $b$ are chosen in such a way that $\bar{m}=a/(a+b)=m_{\rm{opt}}$. Keeping the mean fixed at $m_{\rm{opt}}$ we increased the variance $\sigma^{2}$ of the distribution $f$ and looked at $\langle\lambda_{1}\rangle_{m}/\tilde{\lambda}_{1}(\beta_{\rm{opt}}|n,\alpha)$ as a function of the signal to noise ratio $m_{\rm{opt}}/\sigma$ for three different values of $\alpha$ and $n=8$. {\bf{Fig \ref{bet hedging}}C} shows that even when we have a sub-optimal population of anticipators, our model does better than the bet hedging model as long as the signal is  greater than the noise. For a signal to noise ratio $m_{\rm{opt}}/n\simeq 1$ and low values of $\alpha$, the bet hedging model does progressively better. Since it is likely that transition rate $\alpha$ is small for real biological system, it is reasonable to conclude that anticipation is a better alternative to bet-hedging only when signal is dominant over noise i.e. majority of anticipators do not make large errors in anticipating changes in their environment. It is worth remembering however that we have only considered an optimal population of bet-hedgers, all of whom have a reverse transition rate of $\beta_{opt}$.


\section*{Discussion}

Anticipation is the ability of an organism to predict an upcoming environmental transition and prepare itself accordingly. It is ubiquitously observed in plant and animal kingdom and recently it has been found in cyanobacterium as well. We have adopted a two-state model originally proposed by Thattai and Oudenaarden in the context of bet-hedging in fluctuating environment and modified it to incorporate anticipation. Our "anticipators" mounted pre-emptive transitions from a fit to an unfit state a time $m$ before the environment actually changed thereby incurring an immediate evolutionary cost. We showed that for a periodic environment, this cost is effectively over compensated by the long-term benefit. This is due to the fact that transition to a fit phenotype takes on average $\alpha^{-1}$ amount of time. For small $\alpha$ therefore, one has to wait for a long time before it can make a transition and grow. When the time scale of environment change $n$ is roughly comparable to $\alpha^{-1}$, this creates additional trouble as after a transition an organism finds little time to grow before the environment changes again. Therefore, it is clear that any organism will benefit if it can pre-emptively change when $\alpha$ is small and comparable to the switching rate of the environment. In the limit of high switching rate $\alpha$ and a large $n$ on the other hand, we expect that the range over which anticipation is feasible will decrease. In {\bf{Fig \ref{Eigen surface}}}, we see that anticipation leads to an increase in long-term fitness compared to the case where there is no anticipation. We find that for given $\alpha$ and $n$, there is a range $m_{R}$ over which anticipation pay off as well as an optimal time, denoted as $m_{\rm{opt}}$, when fitness due to anticipation is maximum. Both $m_{\rm{opt}}$ and $m_{R}$ are decreasing functions of $\alpha$ and $n$ ({\bf{Fig \ref{m optimal}}}) for reasons mentioned above, with $m_{\rm{opt}}/n$ decaying as $1/n$ as the number of cell division events $n$ increases. Our results indicate that having a capacity to predict changes in environment increases an organism's over-all chance of survival when the environment shuttles periodically between two states.   

Anticipation can be achieved by maintaining an internal oscillator that is entrained with the periodic environment as is seen in circadian rhythm. These oscillators also exhibit a free-running rhythm in absence of any external reinforcement with a period similar to that of their drivers. Typically, the organism invests in maintaining a sensory apparatus that constantly surveys and samples the environment. Any outside change can then be relayed to the inside by means of a signal transduction network. In light induced entrainment in zebra fish for example, photoreceptors engage the light entrainment pathway \cite{vatine2009light},\cite{whitmore2000light}. Two different sources of noise can be recognized in this context. First, owing to the inherent probabilistic nature of reaction-diffusion processes, outcomes of signal transduction networks are themselves susceptible to stochastic fluctuations. Consequently, the mechanism by which an external signal is coupled to an organism is noisy. Therefore, it stands to reason to look at a distribution of "anticipators", each one of them equipped with a device that is error prone. We looked at an ensemble of such "anticipators" where an error prone anticipator makes a faulty judgment about the environment and mounts a pre-emptive response not at $m_{\rm{opt}}$, but at a time $m- m_{\rm{opt}} \left(m_{\rm{opt}}-m\right)$ after (before) the optimal time. We further assumed that the ensemble average of error is zero. We find that not only does our model increase fitness due to anticipation, it fares better than the two-state bet-hedging model for a wide range of signal to noise ratio ({\bf{Fig \ref{bet hedging}}}), with the latter overtaking our model only when signal to noise ratio is $\sim 1$. 

Secondly, the external driver itself can be stochastic. It is likely that length of duration of a particular environmental state is itself a random variable and hence should in principle be drawn from a distribution. Previously, Jenssen has shown that a driven non-linear oscillator can be synchronized to an external random driver in the limit of strong coupling and when the random driving frequency is not too far off the natural frequency of oscillation \cite{jensen2002synchronization}. Similarly, coherence was demonstrated in FitzHugh-Nagumo model for moderate noisy amplitude of a Gaussian delta-correlated external driver \cite{pikovsky1997coherence}. Recently, using a combination of stochastic simulation and synthetic biology, Butzin et al. demonstrated entrainment in synthetic gene oscillator for telegraph noise \cite{butzin2015marching}. Therefore anticipation can still work as a viable strategy even when an external environment varies at random. We have not looked at duration length fluctuations in environment dynamics. Whether our model can enhance long-term fitness in this case, remains to be seen.

Another concern, which we have ignored so far, is the cost of maintaining a sensory device. Production of sensory proteins (mostly surface receptors) is energy consuming and we have not accounted for that energy cost anywhere in our equations. One straightforward way of incorporating this cost is by suppressing the growth rates by an amount $c$, where $c$ is understood as the cost for maintaining a sensory pathway (see Appendix A2). {\bf{Fig \ref{cost}}} shows the intersection contour of $\langle\lambda_{1}\rangle_{m}/\tilde{\lambda}_{1}(\beta_{\rm{opt}}|n,\alpha)$ surface with $z=1$ plane for two different costs. A contour divides the $n-\alpha$ plane in two halves: upper half begin the region where anticipation is a better alternative to bet-hedging. As expected, with a rise in cost, fitness gain from anticipation is more than bet-hedging, only for large values of $n$ and $\alpha$. 

In sum, central message of this theoretical work is that for a periodically changing environment, an asexually reproducing organism can reap long-term benefit by maintaining a sensory device that enables it to anticipate impending environmental changes. Anticipation is worthwhile provided the duration of any given environmental state is not long enough so that the organism can adapt and the rate of transition from the unfit to the fit phenotype is not so large that a major fraction of the population can change phenotype over one cell division time-scale. Furthermore anticipation can serve as an effective alternate strategy to bet-hedging only when the variance in anticipation is small compared to the mean.

\medskip
\bibliographystyle{unsrt}
\bibliography{Reference_persistence}

\begin{thebibliography}{10}

\bibitem{cohen1966optimizing}
Dan Cohen.
\newblock Optimizing reproduction in a randomly varying environment.
\newblock {\em Journal of theoretical biology}, 12(1):119--129, 1966.

\bibitem{cohen1967optimizing}
Dan Cohen.
\newblock Optimizing reproduction in a randomly varying environment when a
  correlation may exist between the conditions at the time a choice has to be
  made and the subsequent outcome.
\newblock {\em Journal of Theoretical Biology}, 16(1):1--14, 1967.

\bibitem{hobby1942observations}
Gladys~L Hobby, Karl Meyer, and Eleanor Chaffee.
\newblock Observations on the mechanism of action of penicillin.
\newblock {\em Proceedings of the Society for Experimental Biology and
  Medicine}, 50(2):281--285, 1942.

\bibitem{bioger1944treatment}
JW~Bioger.
\newblock Treatment of staphylococcal infections with penicillin by
  intermittent sterilization.
\newblock {\em Lancet Oct}, 14:497, 1944.

\bibitem{balaban2004bacterial}
Nathalie~Q Balaban, Jack Merrin, Remy Chait, Lukasz Kowalik, and Stanislas
  Leibler.
\newblock Bacterial persistence as a phenotypic switch.
\newblock {\em Science}, 305(5690):1622--1625, 2004.

\bibitem{arkin1998stochastic}
Adam Arkin, John Ross, and Harley~H McAdams.
\newblock Stochastic kinetic analysis of developmental pathway bifurcation in
  phage $\lambda$-infected escherichia coli cells.
\newblock {\em Genetics}, 149(4):1633--1648, 1998.

\bibitem{novick1957enzyme}
Aaron Novick and Milton Weiner.
\newblock Enzyme induction as an all-or-none phenomenon.
\newblock {\em Proceedings of the National Academy of Sciences},
  43(7):553--566, 1957.

\bibitem{acar2005enhancement}
Murat Acar, Attila Becskei, and Alexander van Oudenaarden.
\newblock Enhancement of cellular memory by reducing stochastic transitions.
\newblock {\em Nature}, 435(7039):228, 2005.

\bibitem{gardner2000construction}
Timothy~S Gardner, Charles~R Cantor, and James~J Collins.
\newblock Construction of a genetic toggle switch in escherichia coli.
\newblock {\em Nature}, 403(6767):339, 2000.

\bibitem{isaacs2003prediction}
Farren~J Isaacs, Jeff Hasty, Charles~R Cantor, and James~J Collins.
\newblock Prediction and measurement of an autoregulatory genetic module.
\newblock {\em Proceedings of the National Academy of Sciences},
  100(13):7714--7719, 2003.

\bibitem{ozbudak2004multistability}
Ertugrul~M Ozbudak, Mukund Thattai, Han~N Lim, Boris~I Shraiman, and Alexander
  Van~Oudenaarden.
\newblock Multistability in the lactose utilization network of escherichia
  coli.
\newblock {\em Nature}, 427(6976):737, 2004.

\bibitem{Strogatz1994}
S.~H. Strogatz.
\newblock {\em Nonlinear Dynamics and Chaos}.
\newblock Perseus Books, Reading, MA, 1985.

\bibitem{acar2008stochastic}
Murat Acar, Jerome~T Mettetal, and Alexander Van~Oudenaarden.
\newblock Stochastic switching as a survival strategy in fluctuating
  environments.
\newblock {\em Nature genetics}, 40(4):471, 2008.

\bibitem{rosen2012anticipatory}
Robert Rosen.
\newblock Anticipatory systems: Philosophical, mathematical, and methodological
  foundations, 2nd edn.(with contributions by judith rosen, john j. klineman
  and mihai nadin), 2012.

\bibitem{yamazaki2000resetting}
Shin Yamazaki, Rika Numano, Michikazu Abe, Akiko Hida, Ri-ichi Takahashi,
  Masatsugu Ueda, Gene~D Block, Yoshiyuki Sakaki, Michael Menaker, and Hajime
  Tei.
\newblock Resetting central and peripheral circadian oscillators in transgenic
  rats.
\newblock {\em Science}, 288(5466):682--685, 2000.

\bibitem{deguchi1979circadian}
Takeo Deguchi.
\newblock A circadian oscillator in cultured cells of chicken pineal gland.
\newblock {\em Nature}, 282(5734):94, 1979.

\bibitem{plautz1997independent}
Jeffrey~D Plautz, Maki Kaneko, Jeffrey~C Hall, and Steve~A Kay.
\newblock Independent photoreceptive circadian clocks throughout drosophila.
\newblock {\em Science}, 278(5343):1632--1635, 1997.

\bibitem{giebultowicz2000transplanted}
Jadwiga~M Giebultowicz, Ralf Stanewsky, Jeffrey~C Hall, and David~M Hege.
\newblock Transplanted drosophila excretory tubules maintain circadian clock
  cycling out of phase with the host.
\newblock {\em Current Biology}, 10(2):107--110, 2000.

\bibitem{lakin2004circadian}
Patricia~L Lakin-Thomas and Stuart Brody.
\newblock Circadian rhythms in microorganisms: new complexities.
\newblock {\em Annu. Rev. Microbiol.}, 58:489--519, 2004.

\bibitem{cohen2015circadian}
Susan~E Cohen and Susan~S Golden.
\newblock Circadian rhythms in cyanobacteria.
\newblock {\em Microbiology and Molecular Biology Reviews}, 79(4):373--385,
  2015.

\bibitem{tagkopoulos2008predictive}
Ilias Tagkopoulos, Yir-Chung Liu, and Saeed Tavazoie.
\newblock Predictive behavior within microbial genetic networks.
\newblock {\em science}, 320(5881):1313--1317, 2008.

\bibitem{kussell2005phenotypic}
Edo Kussell and Stanislas Leibler.
\newblock Phenotypic diversity, population growth, and information in
  fluctuating environments.
\newblock {\em Science}, 309(5743):2075--2078, 2005.

\bibitem{thattai2004stochastic}
Mukund Thattai and Alexander Van~Oudenaarden.
\newblock Stochastic gene expression in fluctuating environments.
\newblock {\em Genetics}, 167(1):523--530, 2004.

\bibitem{kussell2005bacterial}
Edo~L Kussell, Roy Kishony, Nathalie~Q Balaban, and Stanislas Leibler.
\newblock Bacterial persistence: a model of survival in changing environments.
\newblock {\em Genetics}, 2005.

\bibitem{lachmann1996inheritance}
Michael Lachmann and Eva Jablonka.
\newblock The inheritance of phenotypes: an adaptation to fluctuating
  environments.
\newblock {\em Journal of theoretical biology}, 181(1):1--9, 1996.

\bibitem{vatine2009light}
Gad Vatine, Daniela Vallone, Lior Appelbaum, Philipp Mracek, Zohar Ben-Moshe,
  Kajori Lahiri, Yoav Gothilf, and Nicholas~S Foulkes.
\newblock Light directs zebrafish period2 expression via conserved d and e
  boxes.
\newblock {\em PLoS biology}, 7(10):e1000223, 2009.

\bibitem{whitmore2000light}
David Whitmore, Nicholas~S Foulkes, and Paolo Sassone-Corsi.
\newblock Light acts directly on organs and cells in culture to set the
  vertebrate circadian clock.
\newblock {\em Nature}, 404(6773):87, 2000.

\bibitem{jensen2002synchronization}
RV~Jensen.
\newblock Synchronization of driven nonlinear oscillators.
\newblock {\em American Journal of Physics}, 70(6):607--619, 2002.

\bibitem{pikovsky1997coherence}
Arkady~S Pikovsky and J{\"u}rgen Kurths.
\newblock Coherence resonance in a noise-driven excitable system.
\newblock {\em Physical Review Letters}, 78(5):775, 1997.

\bibitem{butzin2015marching}
Nicholas~C Butzin, Philip Hochendoner, Curtis~T Ogle, Paul Hill, and William~H
  Mather.
\newblock Marching along to an offbeat drum: Entrainment of synthetic gene
  oscillators by a noisy stimulus.
\newblock {\em ACS synthetic biology}, 5(2):146--153, 2015.

\end{thebibliography}
\numberwithin{equation}{subsection}
\section*{Appendix}
\subsection*{A1: Analytical approximation of $\lambda_{1}$}\label{A1}
\renewcommand{\theequation}{A1.\arabic{equation}}
\renewcommand{\thefigure}{S\arabic{figure}}
\setcounter{figure}{0}
From Eqn (\ref{Final State}) we see that the transfer matrix $\mathbb{T}=\mathbb{M}_{4}^{m}\mathbb{M}_{3}^{n-m}\mathbb{M}_{2}^{m}\mathbb{M}_{1}^{n-m}$, where $\mathbb{M}_{1}=\begin{pmatrix} 2 & \alpha \\ 0 & 1-\alpha \end{pmatrix}$, $\mathbb{M}_{2}=\begin{pmatrix} 2-\alpha & 0 \\ \alpha & 1 \end{pmatrix}$, $\mathbb{M}_{3}=\begin{pmatrix} 1-\alpha & 0 \\ \alpha & 2 \end{pmatrix}$ and $\mathbb{M}_{4}= \begin{pmatrix} 1 & \alpha \\ 0 & 2-\alpha \end{pmatrix}$.
We note that 
\begin{eqnarray}
|\mathbb{M}_{1}| & = & |\mathbb{M}_{3}|=2(1-\alpha)\qquad\text{and}\nonumber\\
|\mathbb{M}_{2}| & = & |\mathbb{M}_{4}|=2-\alpha\label{M1M2_determinant}
\end{eqnarray}
where $|..|$ stands for determinant of a matrix. Therefore the determinant of $\mathbb{T}$, denoted as $\Delta$ is given by
\begin{eqnarray}
\Delta&=&|\mathbb{M}_{1}|^{2(n-m)}|\mathbb{M}_{2}|^{2m}\nonumber\\
&=&2^{2(n-m)}(1-\alpha)^{2(n-m)}(2-\alpha)^{2m}\label{determinant}
\end{eqnarray}
With some work, $\eta=\text{Tr}\left(\mathbb{T}\right)$ can be shown to be
\begin{eqnarray}
\eta&=&2\sqrt{\Delta}+\frac{\alpha^{2}}{2^{2m}(1-\alpha)^{2m}(1-\alpha^{2})^{2}}\left((4-2\alpha)^{m}(1-\alpha)^{n+1}+2^{n}(1-\alpha)^{m}(1+\alpha-2(2-\alpha)^{m})\right)^{2}\nonumber\\
&=&2\sqrt{\Delta}+F(n,m,\alpha)\label{trace}
\end{eqnarray}
We will now break the entire range of $\alpha$ into three regions, i) $R_{I}$ ($\alpha\simeq 1$ ), ii) $R_{II}$ ($\alpha$ moderate) and iii) $R_{III}$ ($\alpha\ll 1)$.

\begin{figure}
\begin{center} 
\includegraphics[width=12cm,height=10cm]{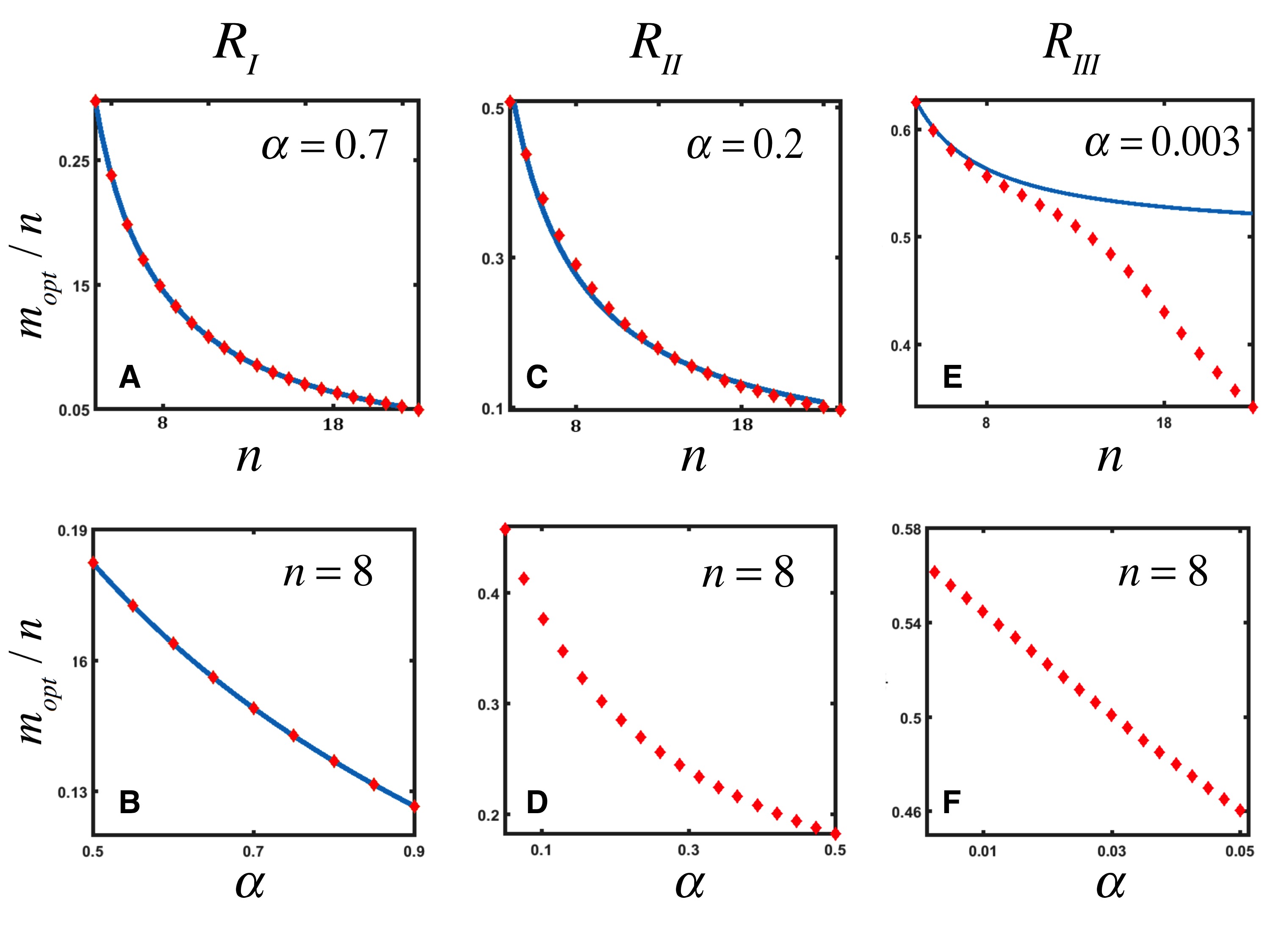}
\caption{{\bf{Plots of $m_{\rm{opt}/n}$ for three different ranges of $\alpha$}}. {\bf{(A-B)}} Plots of $m_{\rm{opt}}/n$ as functions of $n$ and $\alpha$ for $\alpha=0.7$ and $n=8$ respectively. $\alpha$ value used in {\bf{(A)}} and the range of $\alpha$ values chosen for {\bf{(B)}} lie in region $R_{I}$. Numerical calculation is shown in filled red diamond symbols and the analytical approximation obtained from Eqn (\ref{mbest_alpha_large}) is shown in solid blue line. {\bf{(C-D)}} Same as {\bf{(A-B)}}, but the $\alpha$ values are chosen from region $R_{II}$. {\bf{(C)}} is generated using an $\alpha$ value of $0.2$. The blue line in {\bf{(C)}} is the best fit of the numerical data (red diamonds) to a function of the form $a+b/n$. {\bf{(E-F)}} Same as {\bf{(A-B)}}, but $\alpha$ values are taken from $R_{III}$. The solid blue line in {\bf{(E)}} is the plot of Eqn (\ref{mbest alpha small}). {\bf{(E)}} is generated for an $\alpha$ value of $0.003$. Because $m_{\rm{opt}}/n$ is independent of $\alpha$ in Eqn (\ref{mbest alpha small}), we have not shown it in {\bf{(F)}}.} 
\label{m_best all plots}
\end{center}
\end{figure}

\begin{figure}
\begin{center} 
\includegraphics[width=12cm,height=10cm]{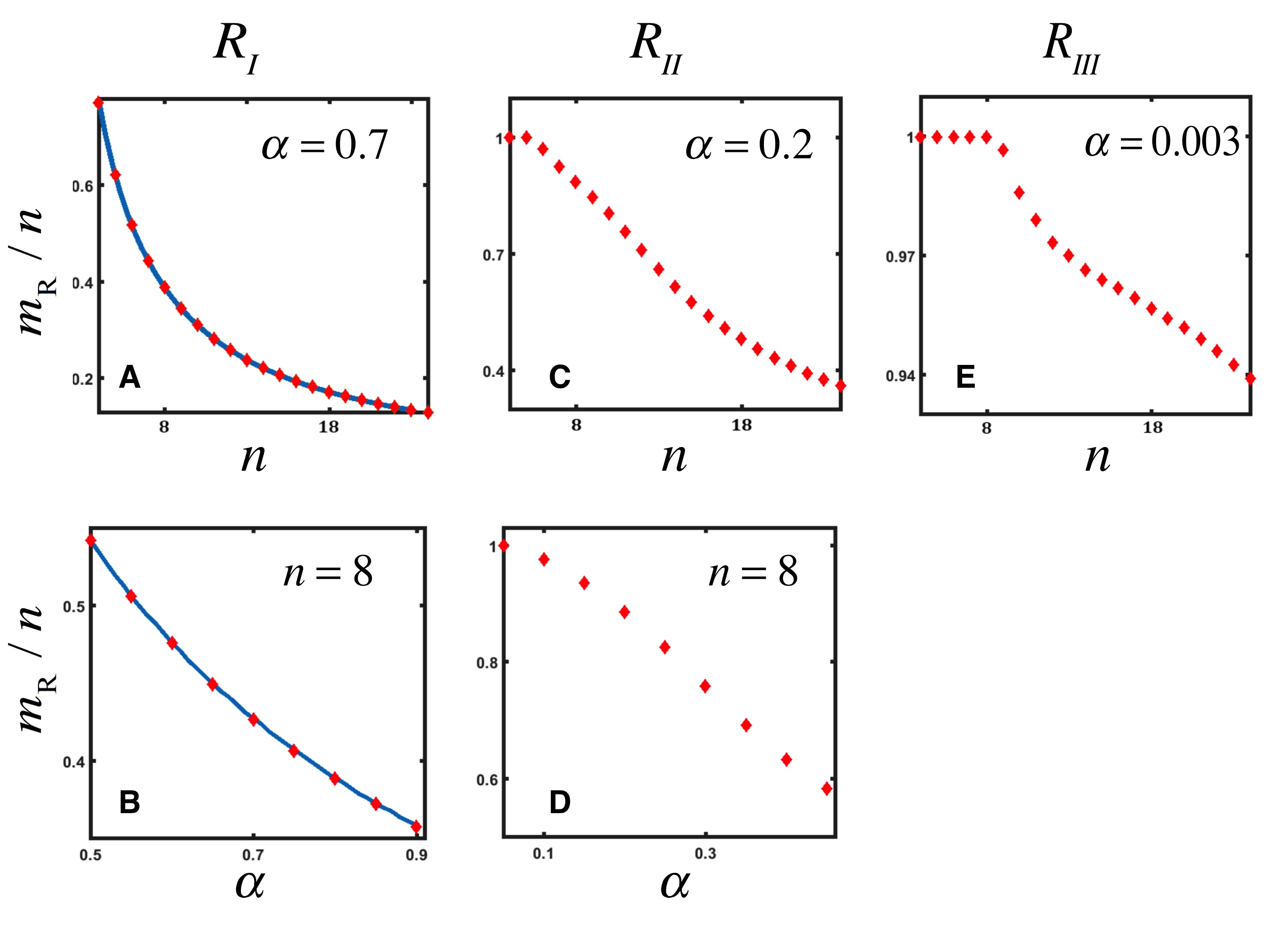}
\caption{{\bf{Plots of $m_{R}/n$ for three different ranges of $\alpha$}}. {\bf{(A-B)}} Plots of $m_{R}/n$ as functions of $n$ and $\alpha$ for $\alpha=0.7$ and $n=8$ respectively. $\alpha$ value used in {\bf{(A)}} and the range of $\alpha$ values chosen for {\bf{(B)}} lie in region $R_{I}$. Numerical calculation is shown in filled red diamond symbols and $m_{R}/n$ obtained by solving the transcendental equation Eqn (\ref{transcendental system mr}) is shown in solid blue line. {\bf{(C-D)}} Same as {\bf{(A-B)}}, but the $\alpha$ values are chosen from region $R_{II}$. {\bf{(C)}} is generated using an $\alpha$ value of $0.2$. {\bf{(E)}} Same as {\bf{(A)}}, but $\alpha$ value used is $0.003$. In region $R_{III}$, typically $m_{R}>n$, which we have manually reset back to $n$. Therefore variation of $m_{R}/n$ as a function of $\alpha$ is not shown in region $R_{III}$.} 
\label{m_range all plots}
\end{center}
\end{figure}

In $R_{I}$, for $m<n$, from Eqn (\ref{determinant}) we have $\Delta\simeq 0$. Therefore the largest eigen value of $\mathbb{T}$ is given by
\begin{eqnarray}
\lambda_{1}\simeq\eta&=&\frac{\alpha^{2}}{2^{2m}(1+\alpha)^{2}}\left(2^{m}(1+\xi)^{m}\xi^{n-m}+2^{n}\frac{(1+\alpha-2(2-\alpha)^{m})}{\xi}\right)^{2}\nonumber\\
&\simeq&\frac{2^{2(n-m)}\alpha^{2}}{(1-\alpha^{2})^{2}}\left(1+\alpha-2(2-\alpha)^{m}\right)^{2}\label{large_lambda_eigen}
\end{eqnarray}
where $\xi=1-\alpha\ll1$. The optimal value of $m$, for which $\lambda_{1}$ reaches a maximum is given by
\begin{eqnarray}
\partial_{m}\lambda_{1}&=&-\kappa ce^{-cm}\left(b-2e^{\delta m}\right)^{2}+4\kappa\delta e^{-cm}e^{\delta m}(2e^{\delta m}-b)\nonumber\label{deivative_large_alpha}=0
\end{eqnarray}
where $\kappa=2^{2n}\alpha^{2}/(1-\alpha^{2})^{2}$, $b=1+\alpha$, $c=2\ln2$ and $\delta=\ln\left(2-\alpha\right)$. Therefore the optimal $m$, $m_{\rm{opt}}$, is given by
\begin{equation}
m_{\rm{opt}}=\frac{1}{\ln(2-\alpha)}\ln\frac{(1+\alpha)\ln2}{2\ln2-2\ln(2-\alpha)}\label{mbest_alpha_large}
\end{equation}
We see that $m_{\rm{opt}}$ is independent of $n$. Therefore when scaled, $m_{\rm{opt}}/n$ will fall off as $1/n$ with an increase in cell division events $n$. The range $m_{R}$ can be calculated by solving the transcendental equation
\begin{eqnarray}
\lambda_{1}\left(m_{R}\right)&=&\lambda_{1}\left(0\right)\qquad\text{yielding}\nonumber\\
2e^{m_{R}\delta}-b&=&\xi e^{m_{R}\ln2}\label{transcendental system mr}
\end{eqnarray}
The plots of $m_{\rm{opt}}/n$ and $m_{R}/n$ as functions of $n$ and $\alpha$, shown in {\bf{Fig \ref{m_best all plots}.A-B}} \& {\bf{Fig \ref{m_range all plots}.A-B}}, demonstrate that Eqn (\ref{mbest_alpha_large}) and Eqn (\ref{transcendental system mr}) agree very well with numerical results.

In region $R_{III}$, approximation to Eqn (\ref{determinant}) and Eqn (\ref{trace}) can be obtained in powers of $\alpha\ll 1$.  Expanding $\sqrt{\Delta}$ in Eqn (\ref{trace}) to $O(\alpha^{2})$ we have
\begin{eqnarray}
\sqrt{\Delta}&=&2^{n-m}(1-\alpha)^{n-m}(2-\alpha)^{m}\nonumber\\
&=&2^{n}-2^{n-1}(2n-m)\alpha+2^{n-3}\alpha^{2}\left(4(n-m)(n-1)+m(m-1)\right)+O(\alpha^{3})\nonumber\\
&=&\Delta_{1}+\Delta_{2}\alpha+\Delta_{3}\alpha^{2}+O(\alpha^{3})\label{determinant order two}
\end{eqnarray}
Similarly, $F$ in Eqn (\ref{trace}) can be written as $F_{1}\alpha^{2}+F_{2}\alpha^{3}+F_{3}\alpha^{4}+O(\alpha^{5})$. Therefore the $\sqrt{\eta^{2}-4\Delta}$ can be written as 
\begin{eqnarray}
\sqrt{\eta^{2}-4\Delta}&\simeq&\left[4(\Delta_{1}+\Delta_{2}\alpha+\Delta_{3}\alpha^{2})(F_{1}\alpha^{2}+F_{2}\alpha^{3}+F_{3}\alpha^{4})+F_{1}^{2}\alpha^{4}\right]^{\frac{1}{2}}\nonumber\\
&=&2\alpha\sqrt{\Delta_{1}F_{1}}+\alpha^{2}\frac{\Delta_{1}F_{2}+F_{1}\Delta_{2}}{\sqrt{\Delta_{1}F_{1}}}+O(\alpha^{3})\label{second order approx sq root}
\end{eqnarray}
where $F_{1}$ and $F_{2}$ are given by
\begin{eqnarray}
F_{1}&=&2^{-2m}\left(2^{2m}-2^{n}(2^{m+1}-1)\right)^{2}\nonumber\\
F_{2}&=&2^{-2m}\left(2^{2m}-2^{n}(2^{m+1}-1)\right)\times\nonumber\\
&&\left(2^{n+1}+m2^{2m}+m2^{n+m+1}-n2^{2m+1}-2^{2m+1}\right)\nonumber\label{F second order}
\end{eqnarray}
Therefore, 
\begin{eqnarray}
\hspace{-1.0cm}
\lambda_{1}&=&\Delta_{1}+\left(\Delta_{2}+\sqrt{\Delta_{1}F_{1}}\right)\alpha+\left(\Delta_{3}+\frac{1}{2}F_{1}+\frac{\Delta_{1}F_{2}+F_{1}\Delta_{2}}{2\sqrt{\Delta_{1}F_{1}}}\right)\alpha^{2}\label{eigen second approx}
\end{eqnarray}
Keeping up to linear order in $\alpha$, and setting the derivative of Eqn (\ref{eigen second approx}) w.r.t $m$ equal to zero we get
\begin{eqnarray}
\partial_{m}\lambda_{1}&=&2^{n-1}\alpha-ae^{-am}2^{\frac{n}{2}}\alpha\left(e^{2am}-2^{n}\right)=0\nonumber\\
\Rightarrow m_{\rm{opt}}&=&\frac{n}{2}+\frac{1}{a}\ln\left[\frac{1}{4a}+\frac{\sqrt{1+16a^{2}}}{4a}\right]\label{mbest alpha small}
\end{eqnarray}
where $a=\ln2$. Eqn (\ref{mbest alpha small}) is independent of $\alpha$ because of the linear approximation. From Eqn (\ref{mbest alpha small}), we see that $m_{\rm{opt}}/n$, falls of as $1/n$ when $n$ is increased and approaches $1/2$ in the limit of large $n$. This approximation gets progressively worse for large values of $n$ as shown in {\bf{Fig \ref{m_best all plots}.E}}. 

We could not find a suitable approximation to $\lambda_{1}$ in $R_{II}$. But a fit of the form $a+b/n$ for $m_{\rm{opt}}/n$ agrees very well ({\bf{Fig \ref{m_best all plots}.C}}).

\subsection*{A2: Cost of having a sensory apparatus}\label{A2}
\renewcommand{\theequation}{A1.\arabic{equation}}

\begin{figure}
\begin{center} 
\includegraphics[width=8cm,height=8cm]{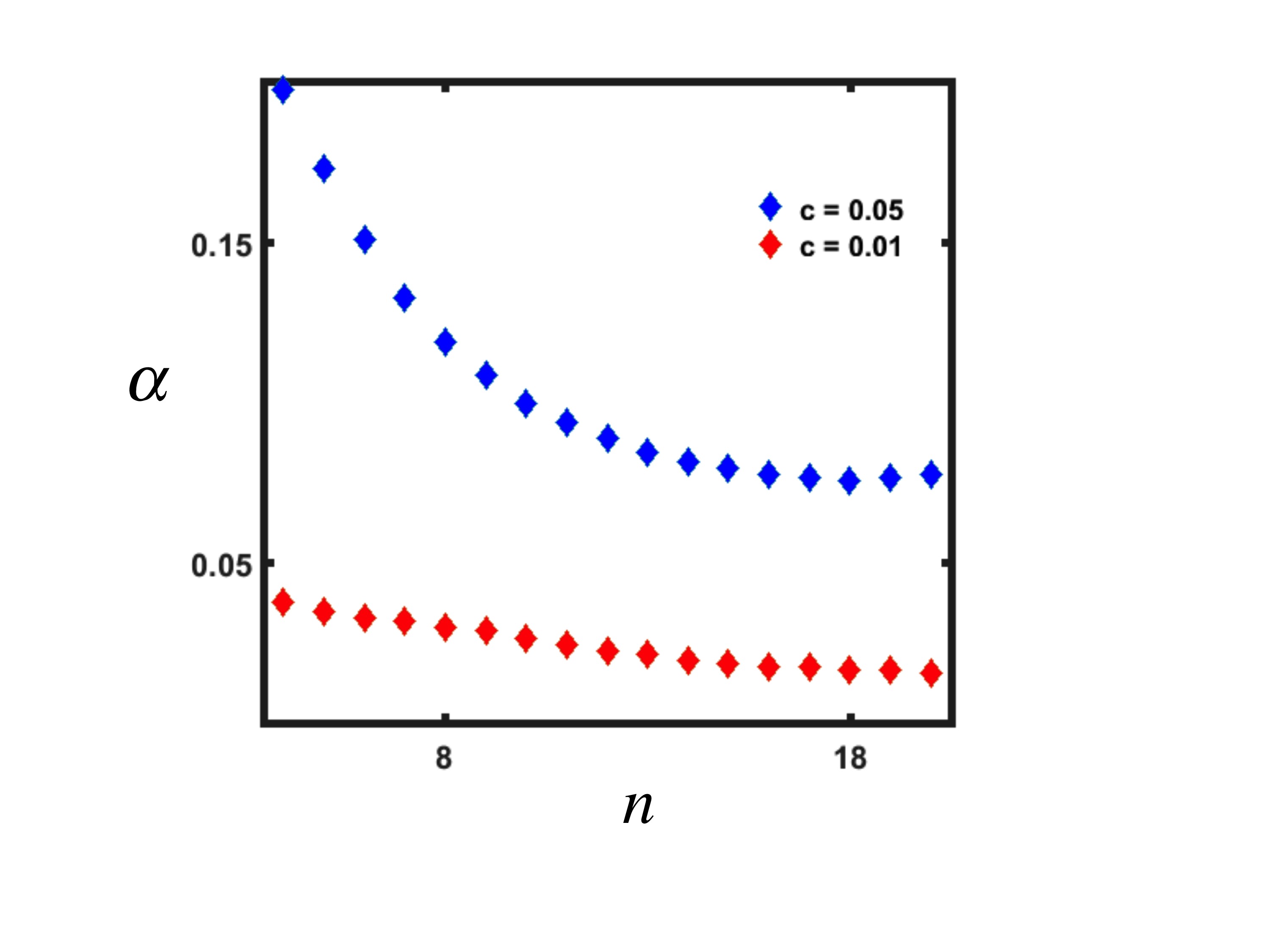}
\caption{Contour of intersection of $\lambda_{1}\left(m_{\rm{opt}}|n,\alpha\right)/\tilde{\lambda}_{1}\left(\beta_{opt}|n,\alpha\right)$ surface with $z=1$ plane for two different costs $c$} 
\label{cost}
\end{center}
\end{figure}

One obvious way of incorporating a cost of having a sensory apparatus is by reducing the growth rate by an amount $c < \gamma$. Defining the dimensionless cost $c=c/\gamma$, the modified matrices are given by $\mathbb{\bar{M}}_{1}=\begin{pmatrix} 2-c & \alpha \\ 0 & 1-\alpha-c \end{pmatrix}$, $\mathbb{\bar{M}}_{2}=\begin{pmatrix} 2-\alpha-c & 0 \\ \alpha & 1-c \end{pmatrix}$, $\mathbb{\bar{M}}_{3}=\begin{pmatrix} 1-\alpha-c & 0 \\ \alpha & 2-c \end{pmatrix}$ and $\mathbb{\bar{M}}_{4}= \begin{pmatrix} 1-c & \alpha \\ 0 & 2-\alpha-c \end{pmatrix}$ respectively. Long-term fitness is given by the largest eigen value $\bar{\lambda}$ of the transfer matrix $\mathbb{\bar{M}}_{1}^{m}\mathbb{\bar{M}}_{3}^{n-m}\mathbb{\bar{M}}_{2}^{m}\mathbb{\bar{M}}_{1}^{n-m}$. {\bf{Fig \ref{cost}}} shows the intersection of $\lambda_{1}\left(m_{\rm{opt}}|n,\alpha\right)/\tilde{\lambda}_{1}\left(\beta_{opt}|n,\alpha\right)$ surface with $z=1$ plane for two different costs. 

\end{document}